%% file: 08-bui-paper.tex
\documentclass[english]{llncs}
\usepackage[
    pdfpagescrop={92 52 523 730},
    colorlinks=true,
    citecolor=black,
    filecolor=black,
    linkcolor=black,
    urlcolor=black,
    breaklinks=true,
    ]{hyperref}

\usepackage{mathtools}
\usepackage{float}

\newcommand{\tuple}[1]{\langle #1\rangle}

\newcommand{\boldtitle}[1]{\vspace{5px}\noindent\textbf{#1}}

\usepackage{prebutterma}
\idline{
Authors' preprint of 2017-08-11.\\Final version to appear in
  \textit{Proceedings of 25th International Workshop on Security Protocols}, Springer LNCS.}

\begin{document}

\title{Key exchange with the help of a public ledger}

\author{Thanh Bui, Tuomas Aura}

\institute{Aalto University \\ \email{thanh.bui@aalto.fi}, \email{tuomas.aura@aalto.fi}}

\maketitle

\pagestyle{plain}
 
\input{08-bui-paper-abstract}

\input{08-bui-paper-intro}
\input{08-bui-paper-background}
\input{08-bui-paper-basic}
\input{08-bui-paper-protocols}
\input{08-bui-paper-applications}

\input{08-bui-paper-discussion}

\input{08-bui-paper-conclusion}

\bibliographystyle{splncs03}

\bibliography{08-bui-paper-references}

\end{document}

%% file: 08-bui-paper-abstract.tex
\begin{abstract}
Blockchains and other public ledger structures promise a new way to create globally consistent event logs and other records. We make use of this consistency property to detect and prevent man-in-the-middle attacks in a key exchange such as Diffie-Hellman or ECDH. Essentially, the MitM attack creates an inconsistency in the world views of the two honest parties, and they can detect it with the help of the ledger. Thus, there is no need for prior knowledge or trusted third parties apart from the distributed ledger. To prevent impersonation attacks, we require user interaction. It appears that, in some applications, the required user interaction is reduced in comparison to other user-assisted key-exchange protocols.  
\end{abstract}

%% file: 08-bui-paper-intro.tex
\section{Introduction}
\label{section:introduction}

Authentication is essential in any key exchange over insecure network. It helps each party be certain that the exchanged key is in fact shared with the desired second party, not an impostor. While authenticated key exchange has been widely studied, it is still a task with many complications. Most of the existing key-exchange protocols require some root of trust between the involved parties, such as prior knowledge on a cryptographic key or password \cite{bellovin1992encrypted,boyko2000provably,diffie1992authentication,hao2010j,jablon1996strong,wu1998secure}, or trusted third-parties (TTP) \cite{dierks2008transport,shamir1984identity}. In situations where these roots of trust are not available, such as device pairing, key exchanges usually rely on user interaction to achieve authentication \cite{alliance2006zigbee,bellovin1992encrypted,gehrmann2004manual,wifispecs,bluetoothspecs,laur2006efficient,vaudenay2005secure}. The probability of a successful attack against these protocols mostly depends on the entropy of the user input. While more complex inputs bring higher security level, they might also cause usability issues.

To ease the requirements of a secure key exchange, we find the solution in blockchains and similar \textit{public ledger} structures, which provide a new way to publish information and achieve consistency in a distributed system. They have been used as key directories that store bindings between identities and public keys \cite{blockstack,kalodner2015empirical,melara2015coniks,yu2015decim}. In a public-key-based key exchange, a party can look up for the public key of the other in these directories. While these systems provide a reasonable user experience and do not depend on any single TTP, they restrict a party to use only the registered keys and thus, complicate key management. Furthermore, they require all interested parties to have a well-defined and unique identity, which does not always hold in reality. For example, two mobile devices of the same user may share the same identity, such as the user's email address, or they may be identified simply by their presence in the user's hands. In any case, the key-identity binding requires the user or someone else to register the bindings. 


In this paper, we present a new family of key exchange protocols that utilizes the global consistency property of the public ledgers. We refer to the protocols as the \textit{public-ledger-based (PLB) protocols}. The idea is to bring transparency to a key exchange by publishing its parameters into the public ledgers. This enables the communication parties to check whether they are participating in the same key exchange, thereby detecting man-in-the-middle (MitM) attacks. Our protocols do not require any roots of trust between the parties or key-identifier bindings. In cases where the parties have public identifiers, such as phone numbers or emails, MitM attacks are prevented without any user interaction. When public identifiers are not available, our protocols still require the users to act as an out-of-band channel. However, the amount of information conveyed via the out-of-band channel is not a function of the desired security level. Thus, high security level can be achieved with little user interaction. Of course, instead of being MitM, the attacker could perform an impersonation attack against one of the parties. We will present application-specific solutions to the attack. 

The rest of the paper is structured as follows. Section~\ref{sec:background} covers the background about public ledgers. Section~\ref{sec:basic} gives an overview of our solution, and Section~\ref{sec:protocols} describes the PLB protocols in detail. Section~\ref{sec:applications} presents a number of practical applications of the protocols. Section~\ref{sec:discussion} discusses the denial-of-service attacks against the protocols and the privacy issues and presents some solutions. Section~\ref{section:conclusion} concludes the paper.

%% file: 08-bui-paper-background.tex

\section{Background}
\label{sec:background}

This section covers the literature about public ledgers and the threat model that we consider in this paper.

\subsection{Public ledger}
\label{sec:public_ledger}
 
The recent widely known compromises of popular service providers \cite{comodohack,diginotarhack,yahoohack} and government surveillances \cite{nsl,fbisecret,applefbi} have motivated quite a few proposals for public ledgers that have no centralized management. These public ledgers are basically public logs of \textit{events} with the following properties \cite{bui2016dac}: \textit{immutability} and \textit{global consistency} of the event history, \textit{inclusiveness} in the sense that all valid data is included. The ledgers might also define a \textit{linear order} or have some built-in concept of time on all data entries.  

Two main architectures have emerged for the public ledgers \cite{chase2016transparency}: (1) totally distributed with no central points of trust or (2) more centralized with one untrusted third party who manages the ledger's content and several trusted auditors who audits its behaviors. 

\boldtitle{Distributed ledgers. }
The most prominent representative of the distributed ledgers is Bitcoin \cite{nakamoto2008bitcoin}. In Bitcoin, an open peer-to-peer (P2P) network with no single party responsible for any critical operation maintains an append-only log of transactions, called \textit{blockchain}. Transactions take place between public keys and are communicated as signed messages in the P2P network. They are mined into blocks approximately every ten minutes. Global consistency is achieved by compressing the global history of transactions into one cryptographic hash value, which can relatively easily be agreed on and communicated to everyone. Inclusiveness and immutability are guaranteed by a novel distributed consensus mechanism and a competitive mining process. 

The Bitcoin's blockchain has become a reliable log for many other applications, such as CommitCoin [4], Factom\footnote{https://www.factom.com/}, Proof of Existence\footnote{https://proofofexistence.com/}, Virtual notary\footnote{https://virtual-notary.org/}, and Tierion\footnote{https://tierion.com/}. It has also inspired hundreds of alternative cryptocurrencies, such as Ethereum \cite{ethereum} and Namecoin \cite{kalodner2015empirical}. A number of central banks and financial institutions have started investing in Bitcoin-based technologies \cite{barrdear2016macroeconomics,canadadigitalcoin,o2013method}.

\boldtitle{Centralized ledgers. }
Although Bitcoin and its variants have achieved a large degree of success, the decentralized nature limits their ability to achieve widespread adoption. The reason is that, for the systems to be completely trustless, users must store the entire blockchain locally and track its progress. Thus, the amount of storage and communication traffic that they require increases linearly with the number of users.

To address the shortcomings of the blockchain, more centralized ledger architectures have been proposed in the literature, in which the ledger content is maintained by an untrusted third party instead of the P2P network. They rely on trusted \textit{auditors} to achieve the desired properties.

Most of the proposals in this category are for monitoring security of the web PKI. Certificate Transparency (CT) \cite{laurie2013rfc} is a public log of all web certificates, which aims to bring transparency and accountability to the CA operations. The log is structured as an \textit{append-only Merkle tree}, in which new records are added to the right of the tree. Accountable Key Infrastructure (AKI) \cite{kim2013accountable} and its kin \cite{basin2014arpki,szalachowski2014policert} are similar solutions, but they deploy an \textit{ordered Merkle tree}, where the data in the leaf nodes is sorted by the domain name instead. Both append-only and ordered Merkle trees enable logarithmic-size proofs of existence and non-existence for certificates, though. In yet another variant of these ideas, PKI Safety Net \cite{szalachowski2016pki} enables verification of both the issuing time order and the non-existence of records for a domain by maintaining two trees that are similar to those in CT and AKI, respectively. Independent \textit{monitors} maintain a copy of the trees and audit their consistency. 

Apart from web PKI, centralized ledger solutions have been suggested for other applications. CONIKS \cite{melara2015coniks} constructs directories of user certificates. DECIM \cite{yu2015decim} keeps track of uses of a public key in a transparency log so that its owner can detect key misuse. Enhanced Certificate Transparency \cite{ryan2014enhanced} extends CT to handle certificate revocation and shows how this extension can be used in end-to-end email or messaging systems. Bui et al. \cite{bui2016dac} apply public ledgers to group management in distributed systems where entities are represented by their public keys and authorization is in the form of signed certificates. 

In this paper, we are not proposing any new ledger structure. Instead, \textit{we assume that a public ledger with the properties mentioned above exists and we use it as an abstract to build our key exchange protocols}.

\subsection{Threat model}
\label{section:threat_model}
We consider a threat model where there are active attackers against key exchanges. The attackers can perform \textit{man-in-the-middle (MitM)} attacks to intercept the communication traffic or \textit{impersonation} attacks to pretend to be one of the involved parties. They can also perform \textit{denial-of-service (DoS)} attacks to cause failures to key exchanges. 

We assume that all the underlying cryptography is secure. Also, the communication channel between any party and the ledger, if exists, is secured under existing security protocols such as TLS/SSL.  

%% file: 08-bui-paper-basic.tex

\section{Basic idea}
\label{sec:basic}

\begin{figure}[h]
	\centering
	\includegraphics[width=0.7\textwidth]{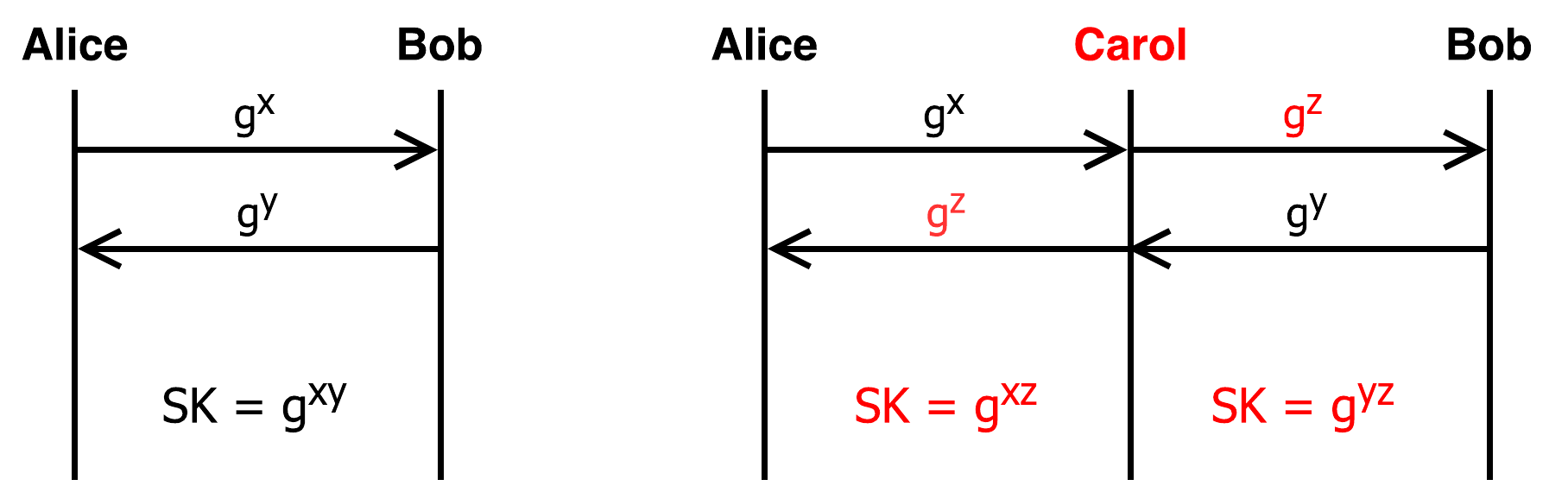}
	\caption{A secure key exchange (on the left) vs a key exchange with MitM attacker (on the right)}
	\label{fig:mitm}
\end{figure}

The key observation that motivated this work is that \textit{in a key exchange that is free of MitM, the endpoints have a consistent view of the key exchange's parameters (i.e. the public keys of the endpoints or the shared key)}. Suppose that when Alice wants to establish a shared key with Bob, Carol performs a MitM attack to intercept the communication. If Carol succeeds, there will be two different key exchanges --- one between Alice and Carol and the other between Carol and Bob, as illustrated in Figure~\ref{fig:mitm}. A secure connection, on the other hand, involves only a single key exchange. In other words, the MitM attack creates inconsistent views of the key exchange between Alice and Bob, while in the case of a secure key exchange, their views are consistent. 

Based on the above observation, \textit{our proposal for detecting MitM attacks against any unauthenticated key exchange is to publish its parameters to a public ledger}. Let us assume such a ledger exists and (unrealistically) that there is only one legitimate key exchange taking place in the world ever. After the unauthenticated key exchange, one of the parties (e.g., the initiator) sends an event $E$ containing the key exchange's data to the ledger, which can be, for example, a hash of the public keys or a hash of the shared key. When the ledger has made the event public, the parties separately query the ledger to check how many new events were entered into the ledger. If there is only one and it matches their view of the key exchange, the key exchange was secure. On the other hand, if two or more events were entered, there may be a MitM attacker around and both parties must discard the results of the key exchange. 


\boldtitle{Context. }
The scheme described above is impractical because it supports only one key exchange ever. To enable multiple key exchanges, our first attempt is to amend the key-exchange event $E$ with a context $C$. The data submitted to the public ledger will be $\tuple{C,E}$ and it is indexed by $C$. That is, one side submits this data to the ledger, both sides query it with the index $C$, and then they compare the received event $E$ (if there is only one) with their view of the key exchange. 

The context can be any information that the two parties agree on \textit{naturally} or \textit{out of band}. The following information is particularly suitable to be used in $C$:

\begin{itemize}
	\item application identifier $a$, 
	\item endpoint identifier $i$ if one is already known to the other party,
	\item user input or user-compared code $c$.
\end{itemize}

The first and second elements are examples of natural contexts, while the last is an example of out-of-band contexts. A key exchange can use both natural context and out-of-band context. How much information the context should hold depends on how many parties are simultaneously writing to the ledger. The frequency of collisions for the index $C$ should be sufficiently small to not frustrate the users. 



\boldtitle{Time window. }
We can see that by tagging each key exchange event with a context, once a context is used, it cannot be re-used for any other key exchange. To optimize the solution, we tag an event $E$ not only with a context $C$ but also with a time $t$. This way, we can do one key exchange per context per \textit{time period} $\tau$. Defining the time period is application-specific, and it must be agreed by the two parties in advance. An application, for example, can define that a context can be used once per minute or per hour.

We can consider time $t$ as a part of the context $C$. However, it must be provided by the ledger because we cannot trust the communication parties to timestamp the key exchange events. Thus, the ledger must have a built-in concept of time. In Bitcoin, for example, a new block appears every ten minutes; thus, we can roughly determine the timestamp of a block by checking how many blocks are after it in the block chain.

\begin{figure}[h]
	\centering
	\includegraphics[width=0.7\textwidth]{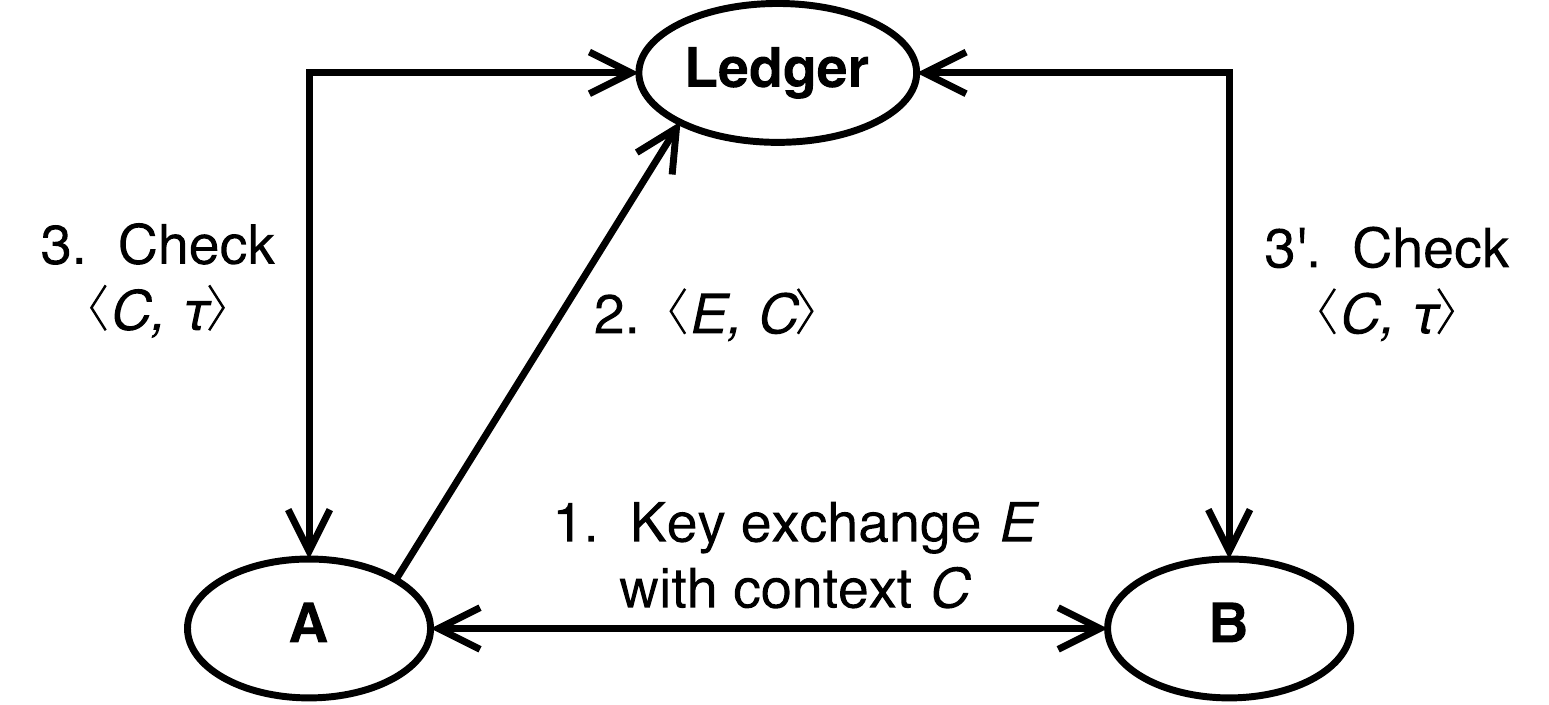}
	\caption{An overview of the public-ledger-based key exchange protocol}
	\label{fig:protocol_overview}
\end{figure}

Figure~\ref{fig:protocol_overview} illustrates an overview of the protocol with the addition of the context and the time period. The data submitted to the ledger will be $\tuple{C,E}$ and it is timestamped by the ledger. When each party checks the ledger, they only query for events with context $C$ that happened in the time period $\tau$. The parties reject the key exchange if there is more than one events that matches this query or if the retrieved event $E$ does not match theirs.



%% file: 08-bui-paper-protocols.tex

\section{Public-ledger-based key exchange}
\label{sec:protocols}

We now describe our PLB key exchange protocols in detail. We will start with a protocol where the communication parties agree on a context out of band and user interaction is required for the agreement. We will then build on it a protocol where the two communication parties must naturally agree on a context with each other and no user interaction is involved.

\subsection{Public-ledger-based key exchange with out-of-band context}
Since the PLB key exchange protocol with out-of-band context requires user interaction, we require the communication parties to be equipped with the following: (1) a screen that can display a string, and (2) a ``start'' button where users can press to trigger an event. 

The protocol involves four parties: the initiator $A$, the responder $B$, the ledger $L$, and the user $U$. It consists of six phases as follows.

\begin{enumerate}
	 \item \textbf{Unauthenticated key exchange}: In the first phase, the parties perform an unauthenticated key exchange to establish a shared secret key $SK$. 



	\item \label{oob_context_acquirement} \textbf{Context acquirement}: The initator $A$ requests the ledger for a context that is not in use. The ledger then returns a context $C$ that matches the query. $A$ then sends the context to $B$ using the shared key. (An alternative way to acquire a context is for one party to generate it locally and try to use it. This way, depending on the entropy of the context, collision might happen, causing failure in the protocol.)


	\item \textbf{Context comparison}: $A$ and $B$ display the context $C$ on their screen so that user $U$ can compare the context. If they display the same context, $U$ presses ``start'' on both endpoints at the same time. After that, both parties immediately query the ledger to get the current ledger's time. Let $S_A$ and $S_B$ be the ledger's times that $A$ and $B$ receive, respectively. 

	\item \textbf{Commitment}: $A$ commits the key exchange to the ledger. It submits an event $E$ containing the key exchange's parameters along with the context $C$ to the ledger. The ledger then stores $E$ and indexes it by $C$. It also gives the event a timestamp $t$. If for some reason (e.g., network failures) the submission fails, $A$ aborts the protocol and informs $B$ about the abortion. Otherwise, $A$ informs $B$ that the commitment was successful via the encrypted channel established in the previous phase. 





			
		

	\item \textbf{Commitment verification}: When the ledger has published the event in the previous phase, both parties separately query the ledger for events with context $C$. $A$ queries for events that happened only in period $[S_A-\triangle, Q_A]$, and $B$ queries for events that happened only in period $[S_B-\triangle, Q_B]$, where $Q_A$ and $Q_B$ are the times when $A$ and $B$ make the query, respectively, and $\triangle$ is an application-specific constant that represents the possible difference in clocks between $A$ and $B$. The ledger then returns a list of events that match the query and a proof that the list is complete. For each event in the list, the ledger also returns the respective timestamp. The parties verify the list. They abort if any of the following occurs: (1) the verification fails, (2) more than one event appear in the list, or (3) the only event in the list does not match their view. Otherwise, the protocol succeeds and the parties accept the shared key. 

	\item \textbf{Confirmation}: User $U$ checks that the protocol succeeds on both sides. It it fails on either side, the user must abort the protocol.






\end{enumerate}



\begin{figure}[h]
	\centering
	\includegraphics[width=0.85\textwidth]{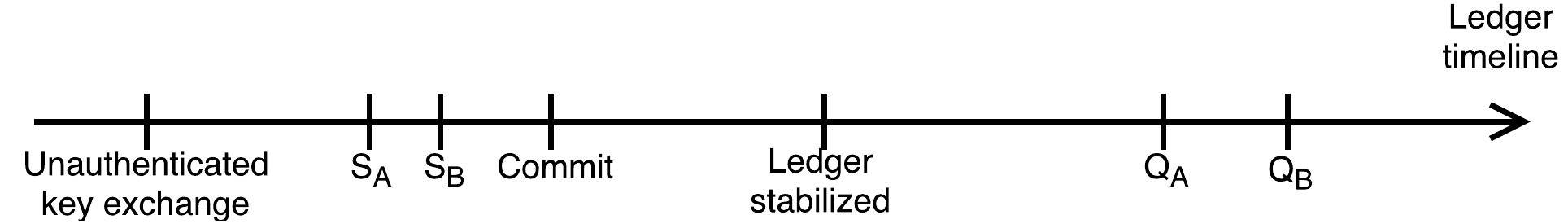}
	\caption{Ledger's timeline of events for a PLB protocol using out-of-band context}
	\label{fig:oob_timeline}
\end{figure}

Figure~\ref{fig:oob_timeline} illustrates all the events in the key exchange above in the ledger's timeline. We can see that if there is man in the middle, the party who queries last ($B$ in the figure) will see that there are two key exchanges with the same context $C$ and thus, abort the protocol. Impersonation attack is also not possible because the user will check that the protocol succeeds on both sides. 

A notable property of the protocol is that its security level does not depend on the entropy of the context. In order for the protocol to succeed, each key exchange simply needs a unique context during its time period. How much information the context should hold depends only on how many parties are simultaneously writing to the ledger.

\subsection{Public-ledger-based key exchange with natural context}
In many cases, the user cannot be present at both sides to press the ``start'' button as in the previous protocol. However, the communication parties in these cases usually have a public identifier that is already known to each other. For example, in an encrypted phone call, the identifiers are the phone numbers or the email addresses. These identifiers can form a context on which the parties can naturally agree with each other. This section will present the PLB key exchange when a natural context is available. This protocol is more simple but requires more assumptions than the previous does because there is no user interaction. 

The PLB key exchange with natural context involves three entities: the initiator $A$, the responder $B$, and the ledger $L$. Beside a context $C$, the initiator and the responder need to agree with each other on the following parameters in advance:

\begin{itemize}
	\item A time period's length $w$, which indicates that a context can only be used once in a time period of length $w$. The first time period of a day starts at 00:00. We assume that the local clocks of the communication parties must be synchronized with reliable sources, and $w$ must be larger than the difference between these clocks.  
	\item A timeout $\alpha$, which indicates the maximum duration of a key exchange. 
\end{itemize}

These parameters can be pre-configured in the communication application. This way, by using the same application to communicate with each other, the two parties naturally agree on the parameters. 

The protocol consists of three phases: \textit{unauthenticated key exchange}, \textit{commitment}, and \textit{commitment verification}. The details of each phase are as follows.

\begin{enumerate}
	 \item \textbf{Unauthenticated key exchange}: As in the previous protocol, the parties first perform an unauthenticated key exchange to establish a shared secret key $SK$. Suppose that the current time period is $[t_0, t_0+w]$ and the current time is $t$. The parties can only perform the unauthenticated key exchange if $t$ is not within $[t_0, t_0+\alpha]$ or $[t_0+w-\alpha, t_0+w]$. This is to ensure that the key exchange does not span over two time periods. Let $S_A$ and $S_B$ be the times when the key exchange starts on $A$ and $B$, respectively. $A$ will abort the protocol if it does not finish by time $S_A+\alpha$, and $B$ will do so by time $S_B+\alpha$.

		
	\item \textbf{Commitment}: The initator $A$ submits an event $E$ containing the key exchange's parameters along with the context $C$ to the ledger. The ledger then stores $E$ and indexes it by $C$. It also gives the event a timestamp $t$. If for some reason (e.g., network failures) the submission fails, $A$ aborts the protocol and informs $B$ about the abortion. Otherwise, $A$ informs $B$ that the commitment was successful via the encrypted channel established in the previous phase. 




	\item \textbf{Commitment verification}: When the ledger has published the event in the previous phase, both parties separately query the ledger for all events with context $C$ in the current period. The ledger then returns a list of events that match the query and a proof that the list is complete. For each event in the list, the ledger also returns the respective timestamp. The parties verify the list. They abort if any of the following occurs: (1) the verification fails, (2) more than one event appear in the list, or (3) the only event in the list does not match their view. Otherwise, the protocol succeeds and the parties accept the shared key. 



\end{enumerate}

The protocol described above can prevent MitM attacks as follows. Assuming that there is a MitM attacker, the attacker must perform a separate key exchange with each party. However, the two key exchanges must share the same context $C$. With time synchronization and the timeout, we ensure that the communication parties perform the key exchange in the same time period. Thus, at least one of the parties will see two key exchanges with the same context published in the ledger and abort the protocol. 

We can see that instead of being the MitM, the attacker could perform an impersonation attack against one of the honest parties (and denial-of-service against the other if necessary). Then, one party will be see the expected one key exchange in the ledger while the other will not even check. We will present application-specific solution to detect impersonation in the next section.

ư

%% file: 08-bui-paper-applications.tex
\section{Applications}
\label{sec:applications}
We now consider a number of practical applications of the PLB key exchange protocol.

\subsection{Encrypted VoIP}
We first present the most convincing application of the protocol --- \textit{encrypted voice over IP (VoIP)}, such as Cryptophone\footnote{http://www.cryptophone.de/}, SilentPhone\footnote{https://www.silentcircle.com/}, and Signal\footnote{https://whispersystems.org/}. 

In these applications, each user is typically represented by a public-key pair, and the users use these key pairs to establish end-to-end encrypted phone calls. To verify that they are actually communicating with the parties they intend, the users can rely on trusted authorities who vouch for the users' public keys, such as the certificate authority in the X.509 Public Key Infrastructure. A major issue with trusted authorities is that they can vouch for fraudulent keys in an attack \cite{diginotarhack,comodohack}. 
An alternative approach is that the users perform key verification out of band. A common approach for key verification is to compare key fingerprint, i.e. some representation of a cryptographic hash of the public keys. Key fingerprint comparison introduces severe usability because users have to manually perform key verification before communicating with a new partner. 
Shorter strings can be provided instead to simplify the verification \cite{blossom1999vp1,zimmermann2011zrtp}. A short authentication string (SAS) is a truncated cryptographic hash of the key exchange's parameters, which is usually represented by a human-friendly format. Participants compute the SAS based on the key exchange they observe and compare the resulting value by reading it aloud. The authentication is based on recognition of the other party's voice.

The PLB key exchange protocol can be applied readily to encrypted VoIP and improve the usability. Specifically, the parties perform the protocol with the context being the phone numbers (or user IDs). This context is inherent to both parties; thus, no out-of-band channel is needed for context verification. The use of the public ledger prevents MitM attacks, and the users will naturally start speaking secrets only if they recognize each other's voices. Therefore, the protocol achieves the same level of security as the SAS verification approach's but better usability because no user interaction is required. 

Of course, users who have never heard each other's voice cannot rely on our protocol to establish a secure phone channel (same as with the SAS verification solution). Voice spoofing is also an issue of voice-based authentication \cite{gupta2007security,petraschek2008security}. Thus, we do not aim to completely replace the stronger methods of authentication, such as manual key fingerprint verification, but to complement them. 

There have been attempts to publish bindings between user identities and public keys to public ledgers \cite{blockstack,kalodner2015empirical,melara2015coniks,yu2015decim}. The main difference between these solutions to ours is that they restrict a party to use only the registered keys and thus, complicate key management. Our protocol, on the other hand, allows users to use arbitrary key pair for each communication, making the use of multiple devices with separate keys easier.

\subsection{Device pairing}
Device pairing --- the procedure of establishing a shared key for secure communication between two devices --- is another application of the PLB protocol. In device pairing, there is usually no key management infrastructure or pre-shared secret between the devices. Thus, authentication solely depends on some form of user action, such as entering a passkey \cite{bellovin1992encrypted,gehrmann2004manual,wifispecs} or verifying numeric code \cite{alliance2006zigbee,bluetoothspecs,laur2006efficient,vaudenay2005secure}. The security level of these approaches depend on the entropy of the user input. In Bluetooth, for example, users have to compare six digits displayed on the devices and the probability of a successful attack is $10^{-6}$. Using longer or more complex inputs would be more error-prone and reduce usability. 

The PLB key exchange protocol might offer advantages other the existing solutions in this scenario. The devices can run the PLB key exchange protocol with out-of-band context, and the users compare the context displayed on the devices. If they are equal, the key exchange succeeds with impossibility of MitM attacks, regardless of the entropy of the context.   

The protocol can be applied to group association where more than two devices want to establish secure communication with each other. 




%% file: 08-bui-paper-discussion.tex

\section{Discussion}
\label{sec:discussion}
In this section, we discuss denial-of-service attacks against the PLB protocols as well as their privacy issues. 

\subsection{Denial-of-service}
\label{sec:dos}

We have showed how the PLB key exchange protocols can defend against MitM and impersonation attacks. Another security vector to the protocols is the denial-of-service attacks where the attacker spams events to the ledger. This could prevent key exchanges to succeed because the endpoints could observe several key exchanges with the same context as theirs in the ledger and drop the key exchange as a result. In this section, we will analyze the causes of these attacks and how we can mitigate them.

The PLB key exchanges using natural contexts are attractive targets of DoS attacks. It is due to the fact that natural contexts are usually endpoint identifiers, which are known by everyone. To permanently prevent any pair of parties to communicate, an attacker can determine the context based on their identifiers and continuously send fake events with the context to the ledger. A solution for this type of targeted attacks is for the ledger to authenticate endpoint identifiers. Authentication could be easily adopted by centralized ledgers like Certificate Transparency. Also, the service provider can act as the ledger provider itself; thus, users can use the same credentials for the application and ledger. 

The out-of-band contexts might also be guessed if their entropy is low. Thus, they must be random so that it is difficult to guess by the attackers but not too complex for users to compare. Using the ledger-provided contexts as in our protocol can limit the attack to some extent but cannot prevent massive spamming. Limitation on rate of submissions, posting fee (e.g. the transaction fee in Bitcoin), and proof-of-work of clients (e.g. client puzzels \cite{aura2000resistant}), are some solutions to prevent massive spamming.

The ledger might be the source of context leak. Depending on the ledger's architecture, the attackers can read the contexts immediately when new events are entered the ledger and then post fake events with the same contexts. The blockchain with P2P network is an example of such designs, in which transactions are propagated through the network for a while before they are mined into blocks. Thus, to prevent context leak, the ledger could be designed so that it can only be queried for individual contexts and and the response for any individual query leaks no information about other events in the ledger. CONIKS's \cite{melara2015coniks} is an example of such design. The ledger could also be round-based, meaning that events can be accepted in a round and only published in the next round. 

\subsection{Privacy}
We now argue about the privacy issue of the PLB protocol that uses the communication parties' identifiers as the context. It is true that if a record is published in the ledger every time two parties communicate with each other, the protocol has serious privacy issue. However, that is not how applications work in practice. Two communication parties need to perform a key exchange only once and save the result locally for later uses. They do not have to perform another key exchange unless one of them changes its public key. Authentication also helps to prevent privacy issues. For example, a party can only query the ledger for the records that are associated with contexts containing its identifier. Furthermore, unlike ledgers that store key bindings, the content of our ledger can be erased periodically without affecting any key exchanges. This not only reduces the footprint of the ledger but also might be of little help in preserving user privacy.

%% file: 08-bui-paper-conclusion.tex
\section{Conclusion}
\label{section:conclusion}
We have shown that it is possible to have authenticated key exchange protocols that take advantages of the global consistency property of blockchains and other public ledgers to detect and prevent man-in-the-middle attacks. While our protocol requires the users to act as an out-of-band channel, the amount of information conveyed out of band is not a function of the desired security level. Instead, it depends on the number of simultaneous key exchanges and thus, on the load of the public ledger; the out-of-band information simply ensures each key exchange gets its unique context string. Further work is required on the detailed requirements and design of the public ledger and on analysis of denial-of-service threats. 